\documentclass[aps,prl,twocolumn,10pt,floatfix]{revtex4-2} 
\usepackage{amsmath, amssymb, amsfonts, bm, mathtools}

\usepackage{graphicx}
\usepackage{epsfig}
\usepackage{epstopdf}  

\usepackage{color,xcolor}
\usepackage{multirow}
\usepackage{enumerate}
\usepackage{soul}

\usepackage{siunitx}
\AtBeginDocument{\RenewCommandCopy\qty\SI} 

\usepackage{physics}

\usepackage{hyperref}
\usepackage{etoolbox}
\makeatletter
\patchcmd{\@float}{\addcontentsline}{\phantomsection\addcontentsline}{}{}
\makeatother

\hypersetup{
    colorlinks=true,
    linkcolor=blue,
    citecolor=blue,
    urlcolor=blue,
    pdfborder={0 0 0}
}

\usepackage[T1]{fontenc}

\usepackage{mathrsfs}

\begin{document}


\title[]{Exceptional Point-enhanced Rydberg Atomic Electrometers}

\author{Chao Liang$^{1,2}$}
\email{liangchaos@mail.tsinghua.edu.cn}
\author{Ce Yang$^{3}$}
\author{Wei Huang$^{3}$}
\author{Li You$^{2,4,5,6}$}

\affiliation{$^{1}$State Key Laboratory of Low-Dimensional Quantum Physics, Beijing Tsinghua Institute for Frontier Interdisciplinary Innovation, Beijing 102200, China}
\affiliation{$^{2}$State Key Laboratory of Low-Dimensional Quantum Physics, Department of
Physics, Tsinghua University, Beijing 100084, China}
\affiliation{$^{3}$China Academy of Aerospace System and Innovation, Beijing 100088, China}
\affiliation{$^{4}$Beijing Academy of Quantum Information Sciences, Beijing 100193, China}
\affiliation{$^{5}$Frontier Science Center for Quantum Information, Beijing, China}
\affiliation{$^{6}$Hefei National Laboratory, Hefei, Anhui 230088, China}


\begin{abstract}

Rydberg atoms, with their large transition dipole moments and extreme sensitivity to electric fields, have attracted widespread attention as promising candidates for next-generation quantum precision electrometry. Meanwhile, exceptional points (EPs) in non-Hermitian systems have opened new avenues for ultrasensitive metrology. Despite increasing interest in non-Hermitian physics, EP-enhanced sensitivity has rarely been explored in Rydberg atomic platforms. Here, we provide a new theoretical understanding of Autler-Townes (AT) effect-based Rydberg electrometry under non-Hermitian conditions, showing that dissipation fundamentally modifies the spectral response and enables sensitivity enhancement via EP-induced nonlinearity. Experimentally, we realize a second-order EP in a passive thermal Rydberg system without requiring gain media or cryogenics, and demonstrate the first EP-enhanced atomic electrometer. The EP can be tuned in real time by adjusting laser and microwave parameters, forming a flexible and scalable platform. Near the EP, the system exhibits a square-root response, yielding a nearly 20-fold enhancement in responsivity. Using amplitude-based detection, we achieve a sensitivity of {$22.68(3)~\mathrm{nVcm^{-1}Hz^{-1/2}}$} under realistic conditions. Our work establishes a practical, tunable platform for EP-enhanced sensing and real-time control, with broad implications for quantum metrology in open systems.
\end{abstract}

\maketitle

Rydberg atoms~\cite{Na}---atoms excited to high principal quantum numbers---possess exaggerated properties such as large electric dipole moments, making them highly sensitive probes for electric field detection. Rydberg-atom-based microwave (MW) detection has emerged as a frontier in quantum precision measurement~\cite{Degen2017}, offering transformative capabilities that depart fundamentally from classical electronics. Over the past decade, Rydberg electrometry has achieved major progress in both principle and performance~\cite{Schlossberger2024,Zhang2024d,Yuan2023a,Liu2023d,Artusio-Glimpse2022,Fancher2021}. Early work employed electromagnetically induced transparency (EIT) and Autler-Townes (AT) splitting to detect weak MW fields with high precision~\cite{Holloway2014,Gordon2010,Sedlacek2012b,Sedlacek2013}. Based on this foundation, methods such as superheterodyne detection~\cite{10.1063/1.5088821,Gordon2019,Jing2020a} and many-body enhancement through phase transitions~\cite{Ding2022,Wu2024} have significantly improved sensitivity.

Among these, EIT remains the most widely used technique for reading AT-induced level splitting in Rydberg atoms~\cite{Holloway2014,Gordon2010,Sedlacek2012b,Sedlacek2013,Sedlacek2012b,Cai2022,Mohapatra2008,Mohapatra2007}. In a typical ladder-type EIT setup, a radio-frequency (RF) electric field drives transitions between adjacent Rydberg states, producing observable splitting $\Delta f$ in the probe transmission spectrum. The splitting relates linearly to the field strength $E$ as (set reduced Planck’s constant $\hbar=1$ for simplicity):
\begin{equation}
	\Delta f = \Omega = \mu_\mathrm{d} E,
	\label{eq1}
\end{equation}
where $\Omega$ is the RF Rabi frequency and $\mu_\mathrm{d}$ the transition dipole moment. This direct relationship allows absolute field measurements traceable to the SI.

However, real Rydberg systems are inherently dissipative due to spontaneous emission and decoherence, making the dynamics non-Hermitian. In such regimes, the linear response of Eq.~(\ref{eq1}) fails, especially near exceptional points (EPs)---non-Hermitian degeneracies where eigenvalues and eigenvectors coalesce~\cite{RevModPhys.93.015005,aar7709,Feng2017,El-Ganainy2018,Li2023,Ding2022z,PhysRevLett.113.053604,Heiss_2012}. Near an EP, a small perturbation $\epsilon$ induces splitting that scales as $\Delta f \propto \sqrt{\epsilon}$, enhancing sensitivity by a factor of $1/\sqrt{\epsilon}$. EP-enhanced sensing has been demonstrated in various platforms~\cite{Li18}, ranging from optical cavities~\cite{PhysRevLett.112.203901,Hodaei2017,Chen2017,Mao2024,Ruan2024,Lai2019,Hokmabadi2019,Xu2024}, magnonic systems~\cite{Wang2024}, photonic crystals~\cite{Park2020}, circuit systems~\cite{Xiao2019,Assawaworrarit2017,Kononchuk2022,Dong2019,Chen2018s,Li2023b,Suntharalingam2023,Yang2023c}, quantum system~\cite{Yu2020} ，and atomic systems~\cite{Liang2023}.

Despite progress, the benefits of EP sensing remain under debate~\cite{Langbein2018,Chen2019a,Wang2020a,Ding2023a,Duggan2022a,Wiersig2020,Zhang2019a,Wiersig2020a,PhysRevLett.132.243601,PhysRevResearch.6.023148}, due to concerns such as amplified noise in gain-assisted systems. In contrast, passive non-Hermitian systems---where only loss is present---can exhibit enhanced noise resilience and operational stability~\cite{Dong2025}. For intrinsically lossy platforms like Rydberg atoms, harnessing dissipation constructively to enhance sensor performance remains an open and largely unexplored opportunity.

\begin{figure}[tbp]
	\centering
	\includegraphics[width=\linewidth]{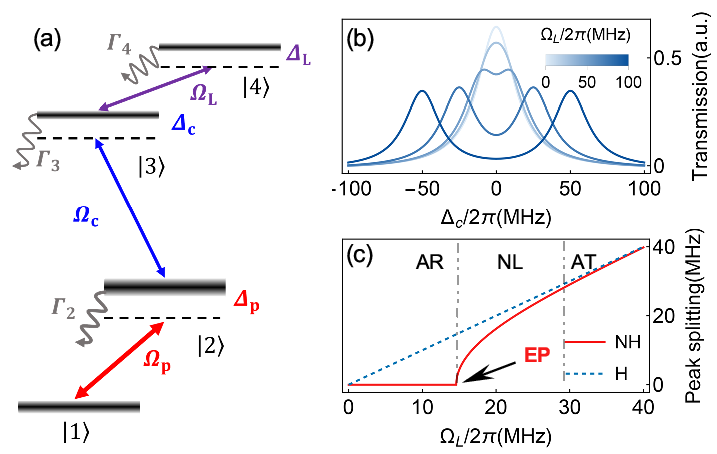}
	\caption{System model for observing EPs in Rydberg atomic ensembles. (a) Energy-level diagram of a four-level Rydberg system driven by probe ($\Omega_{\mathrm{p}}$), coupling ($\Omega_{\mathrm{c}}$), and MW ($\Omega_{\mathrm{L}}$) fields. The probe and coupling lasers form a ladder-type EIT via $|1\rangle \leftrightarrow |2\rangle \leftrightarrow |3\rangle$, while the MW field couples Rydberg states $|3\rangle \leftrightarrow |4\rangle$. Detunings $\Delta_{\mathrm{p}}$, $\Delta_{\mathrm{c}}$, $\Delta_{\mathrm{L}}$ and decay rates $\Gamma_{2,3,4}$ correspond to respective states.
		(b) Probe transmission spectra versus coupling detuning $\Delta_{\mathrm{c}}$ at various MW Rabi frequencies $\Omega_{\mathrm{L}}$. Increasing $\Omega_{\mathrm{L}}$ enhances EIT peak splitting.
		(c) Peak splitting versus $\Omega_{\mathrm{L}}$: the Hermitian case (blue dashed) shows linear scaling, while the non-Hermitian case (red solid) exhibits nonlinear enhancement near the exceptional point (EP). Regions labeled AR, NL, and AT denote absorption, nonlinear, and Autler-Townes regimes, respectively.}
	\label{fig1}
\end{figure}

In this work, we demonstrate an EP-enhanced Rydberg atomic electrometer using a thermal vapor cell. By leveraging the inherent dissipation in the Rydberg excitation manifold and tailoring the coupling fields, we engineer a non-Hermitian Hamiltonian exhibiting a second-order EP. In this regime, the AT splitting becomes intrinsically nonlinear, in excellent agreement with experiment. Near the EP, the sensor’s responsivity to weak microwave (MW) fields is enhanced by nearly a factor of 20. By coupling the signal MW field with a tunable local dressing field, our setup enables amplitude- and phase-sensitive detection. In the nonlinear region, we observe a strongly amplified response, achieving a sensitivity of {$22.68(3)~\mathrm{nVcm^{-1}Hz^{-1/2}}$}. We also detect high-harmonic responses near the EP, suggesting a new mechanism for field control and modulation. These results establish a new paradigm of EP-enhanced quantum electrometry, uniting the intrinsic sensitivity of Rydberg atoms with non-Hermitian criticality.{Compared with recently developed Rydberg electrometry approaches~}\cite{Ding2022,Jing2020a,Gordon2019,Borowka2024}{, our sensitivity is therefore competitive and falls within the advanced range of current leading techniques.} 

As shown in Fig.~\ref{fig1}(a), our system is based on a general four-level Rydberg scheme using two-photon excitation. State $|1\rangle$ is the ground state, $|2\rangle$ is an intermediate excited state with high dissipation $\Gamma_2$, and $|3\rangle$, $|4\rangle$ are Rydberg states with lower decay rates $\Gamma_3,\Gamma_4\ll\Gamma_2$. The probe and coupling lasers drive the $|1\rangle \leftrightarrow |2\rangle$ and $|2\rangle \leftrightarrow |3\rangle$ transitions, with Rabi frequencies and detunings \(\Omega_\mathrm{p}, \Delta_\mathrm{p}\) and \(\Omega_\mathrm{c}, \Delta_\mathrm{c}\), respectively. A local MW field couples $|3\rangle$ and $|4\rangle$ with \(\Omega_\mathrm{L}, \Delta_\mathrm{L}\).

In the rotating frame, the system Hamiltonian reads:
$H = -\Delta_{\mathrm{p}} |2\rangle\langle2|
- (\Delta_{\mathrm{p}} + \Delta_{\mathrm{c}}) |3\rangle\langle3|
- (\Delta_{\mathrm{p}} + \Delta_{\mathrm{c}} + \Delta_{\mathrm{L}}) |4\rangle\langle4| 
+ \frac{1}{2} \left( 
\Omega_{\mathrm{p}} |1\rangle\langle2| 
+ \Omega_{\mathrm{c}} |2\rangle\langle3| 
+ \Omega_{\mathrm{L}} |3\rangle\langle4| 
+ \mathrm{H.c.} 
\right).$
The density matrix $\rho$ evolves under the master equation $\dot{\rho} = -i[H, \rho] + \mathcal{L}[\rho]$, where $\mathcal{L}$ is the Lindblad superoperator incorporating decay and dephasing. When the decay rate $\Gamma_2$ is much greater than all other rates, the intermediate state $|2\rangle$ can be adiabatically eliminated, yielding an effective non-Hermitian Hamiltonian $\mathcal{H}_{\mathrm{NH}}$ for the reduced three-level system~\cite{Liang2023,supp1}. For simplicity, we set all detunings $\Delta_{\mathrm{p,c,L}} = 0$ in the analysis.
\begin{equation}
	\begingroup
	\renewcommand{\arraystretch}{1.3}
	\mathcal{H}_{\mathrm{NH}} = \frac{1}{2}
	\left(
	\begin{array}{ccc}
		-i\gamma_{\mathrm{p}} & -i\Omega_{\mathrm{eff}} & 0 \\
		i\Omega_{\mathrm{eff}} & -i\gamma_{\mathrm{c}} & -\Omega_{\mathrm{L}} \\
		0 & -\Omega_{\mathrm{L}} & 0
	\end{array}
	\right),
	\endgroup
\end{equation}
where $\Omega_{\mathrm{eff}} = \Omega_{\mathrm{p}} \Omega_{\mathrm{c}} / \Gamma_{2}$ is the effective coupling strength between the ground state $\ket{1}$ and the Rydberg state $\ket{3}$. The terms $\gamma_{\mathrm{p(c)}} = \Omega_{\mathrm{p(c)}}^2 / \Gamma_{2}$ represent the effective decay rates induced by the probe field $\Omega_{\mathrm{p}}$ and the coupling field $\Omega_{\mathrm{c}}$, respectively \cite{Liang2023}.  Under the typical condition that the probe field is much weaker than the coupling field, $0 \approx \Omega_{\mathrm{p}} \ll \Omega_{\mathrm{c}}$ (as is the case in most experiments), the three eigenvalues of the non-Hermitian Hamiltonian $\mathcal{H}_{\mathrm{NH}}$ can be approximately expressed as:
\begin{equation}
	E_0 = 0, \quad 
	E_{\pm} = \frac{-i \gamma_{\mathrm{c}} \pm \sqrt{4 \Omega_{\mathrm{L}}^2 - \gamma_{\mathrm{c}}^2}}{4}.
	\label{eq3}
\end{equation}

\begin{figure}[tbp]
	\centering
	\includegraphics[width=\linewidth]{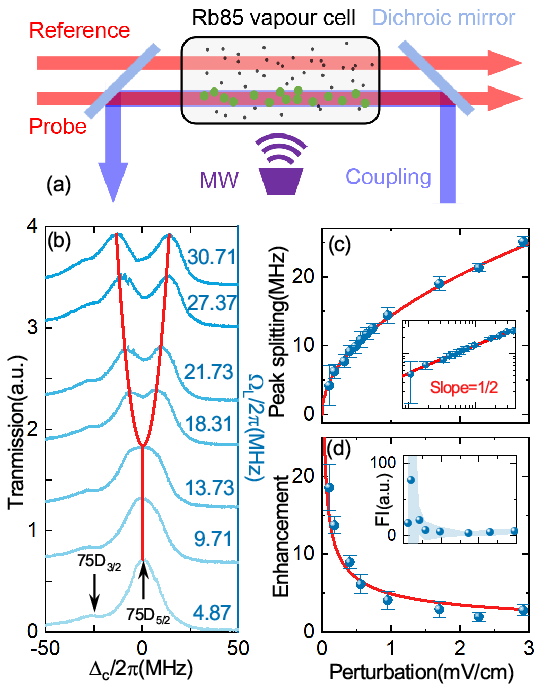}
	\caption{Experimental demonstration of EPs and enhanced response near EPs. (a) Experimental setup: probe and reference beams (red arrows) propagate parallel through a room-temperature $^{85}$Rb vapor cell. The probe counter-propagates with a coupling beam (blue arrow), redirected by a dichroic mirror (DM) to excite Rydberg states. A MW field from a horn antenna couples adjacent Rydberg levels. Transmission difference between probe and reference is detected by a balanced photodetector (PD). (b) Probe transmission spectra versus coupling detuning $\Delta_\mathrm{c}$ for increasing MW Rabi frequencies $\Omega_\mathrm{L}$ (bottom to top). The red curve tracks central peak positions, showing nonlinear splitting near the EP. The main peak corresponds to $5P_{3/2} \leftrightarrow 75D_{5/2}$; the smaller peak to $5P_{3/2} \leftrightarrow 75D_{3/2}$. (c) Peak splitting versus perturbation strength near EP; inset log-log plot with slope 1/2 confirms square-root EP response. (d) Enhancement factor versus perturbation strength;{inset Fisher information versus perturbation strength.} Blue dots: experimental data with error bars (5 measurements); red lines: theoretical fits.}
	\label{fig2}
\end{figure}
As shown in Eq.~(\ref{eq3}), when $\Omega_{\mathrm{L}} = \Omega_{\mathrm{EP}} \equiv \gamma_{\mathrm{c}} / 2$, the real and imaginary parts of the eigenvalues $E_{\pm}$ coalesce, marking a second-order EP. In this regime, the system mimics a $\mathcal{PT}$-symmetric two-level model~\cite{PhysRevLett.127.186601}. For $\Omega_{\mathrm{L}} < \Omega_{\mathrm{EP}}$, the eigenvalues share the same real part but differ in their imaginary components, indicating a $\mathcal{PT}$-broken phase. Conversely, when $\Omega_{\mathrm{L}} > \Omega_{\mathrm{EP}}$, the imaginary parts coincide while the real parts split, characteristic of a $\mathcal{PT}$-symmetry phase.

Near the EP ($\Omega_{\mathrm{L}} \simeq \Omega_{\mathrm{EP}}$), when a small, co-frequency perturbation signal $\Omega_{\mathrm{s}}$ is applied such that $\Omega_{\mathrm{L}} \rightarrow \Omega_{\mathrm{L}} + \Omega_{\mathrm{s}}~( \Omega_{\mathrm{L}}\gg \Omega_{\mathrm{s}})$, the system exhibits a nonlinear response to the electric field, characterized by a square-root scaling of the energy level splitting, i.e., $\Delta f = \mathrm{Re}(E_{+} - E_{-}) \propto \sqrt{\Omega_{\mathrm{s}}}$. In contrast, far from the EP ($\Omega_{\mathrm{L}} \gg \Omega_{\mathrm{EP}}$), the splitting becomes linearly proportional to the perturbation field, $\Delta f = \mathrm{Re}(E_{+} - E_{-}) \propto \Omega_{\mathrm{s}}$, consistent with the conventional AT splitting behavior described in Eq.~(\ref{eq1}) for Hermitian systems. This behavior is confirmed by theoretical simulations, as shown in Fig.~\ref{fig1}(b). With increasing $\Omega_{\mathrm{L}}$, the probe transmission spectrum versus the coupling detuning $\Delta_{\mathrm{c}}$ evolves from a single peak into a clearly split doublet. The peak-to-peak splitting as a function of $\Omega_{\mathrm{L}}$ is plotted in Fig.~\ref{fig1}(c) (red curve), with the Hermitian case indicated by the blue dashed line. Three distinct regimes are identified:  an absorption regime (AR) in the $\mathcal{PT}$-broken phase; and in the $\mathcal{PT}$-symmetry phase, a nonlinear regime (NL) near the EP, followed by the conventional AT regime farther away from the EP.

Based on the theoretical model, we implement the experimental scheme illustrated in Fig.~\ref{fig2}(a). We use a room-temperature vapor cell containing ${}^{85}$Rb and ${}^{87}$Rb. ${}^{85}$Rb atoms are excited to the Rydberg state $\ket{3} = |75D_{5/2}\rangle$ via a two-photon transition through the intermediate state $\ket{2} = |5P_{3/2}, F=4\rangle$, starting from the ground state $\ket{1} = |5S_{1/2}, F=3\rangle$, using counterpropagating 780-nm and 480-nm laser beams. A calcite displacer creates probe/reference beams for differential measurement. A MW field couples $\ket{3}$ and $\ket{4}=|76P_{3/2}\rangle$,{with the corresponding transition dipole moment $\mu_{d} = 3605.9764\, e a_{0}$, where $e$ is the electric charge and $a_{0}$ is the Bohr radius.} Representative experimental results are shown in Fig.~\ref{fig2}(b), where the transmission spectra of the probe field are measured as a function of the coupling detuning $\Delta_{\mathrm{c}}$. From bottom to top, the strength of the MW field  $\Omega_{\mathrm{L}}$ gradually increases, leading to the progressive splitting of the resonance peak. {From the experimental fitting, the Rabi frequency at the EP is found to be $\Omega_L/{2\pi}=14.02$ MHz.}The red line traces the central peak positions, revealing a nonlinear peak-splitting behavior in the vicinity of the EP. 
\begin{figure*}[tbp]
	\centering
	\includegraphics[width=\linewidth]{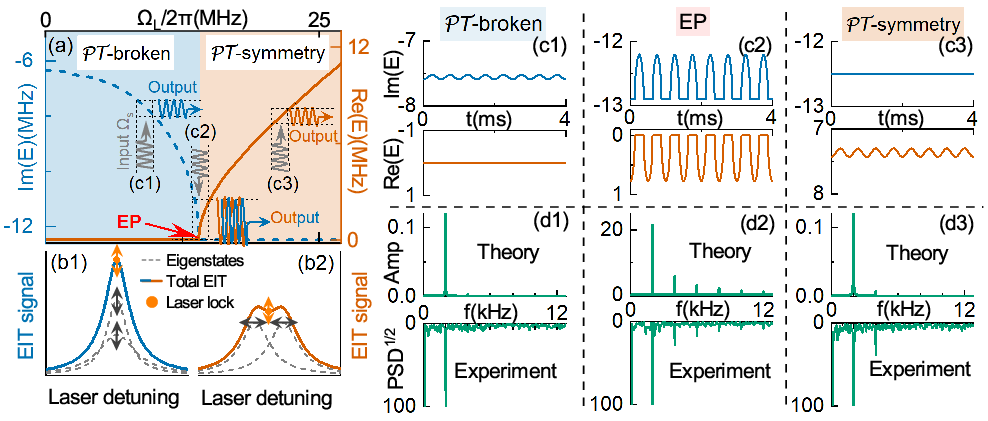}
	\caption{Nonlinear dynamics and signal response near the EP.(a) Complex eigenvalues of the effective non-Hermitian Hamiltonian versus local MW Rabi frequency $\Omega_\mathrm{L}$, showing coalescence of real (orange solid) and imaginary (blue dashed) parts at the EP (red arrow). Blue and orange shaded regions denote the $\mathcal{PT}$-broken and $\mathcal{PT}$-symmetric phases, respectively. Insets show system response under weak signal $\Omega_\mathrm{s}$ (gray line) in three regimes: $\mathcal{PT}$-broken (c1), at EP (c2), and $\mathcal{PT}$-symmetric (c3).(b1,b2) Schematic EIT spectra (blue solid in b1, orange solid in b2) with fitted eigenmodes (gray dashed), illustrating characteristic linewidth and splitting changes of non-Hermitian eigenstates in $\mathcal{PT}$-broken (b1) and $\mathcal{PT}$-symmetric (b2) phases. Weak signal $\Omega_\mathrm{s}$ induces linewidth modulation (b1) or energy-level shifts (b2), causing measurable EIT changes. Locking coupling laser to resonance center makes probe transmission proportional to signal amplitude. (c1-c3) Time evolution of real (top) and imaginary (bottom) parts of eigenvalues under weak signal $\Omega_\mathrm{s}$ in regimes marked in (a). (d1-d3) Fourier spectra of corresponding time-domain responses, with theoretical results (top) and experimental data (bottom). Signal detuning in experiment is $\delta/2\pi=2~\mathrm{kHz}$.}
	\label{fig3}
\end{figure*}
Fig.~\ref{fig2}(c) clearly demonstrates a square-root dependence of the peak splitting on the perturbation electric field near the EP. As shown in the inset of Fig.~\ref{fig2}(c), the slope of $1/2$ in the corresponding log-log plot confirms this characteristic behavior. Owing to the square-root scaling, a significant enhancement in signal response can be achieved compared to the conventional linear scaling. {As shown in Fig.~}\ref{fig2}{(d), our experimental results exhibit an enhancement of nearly 20-fold in responsivity in the vicinity of the EP. Even when accounting for noise amplification, we can achieve an order-of-magnitude improvement in Fisher information (FI) (inset in Fig.~}\ref{fig2}{(d)) in the nonlinear regime~}\cite{supp1}.

In practical scenarios, the frequency of the signal field  $\Omega_{\mathrm{s}}$ is typically unknown and may differ from that of the local field $\Omega_{\mathrm{L}}$. While the enhanced nonlinear energy level splitting near the EP enables the conversion of electric field strength into frequency shifts, this frequency-based readout is ultimately limited by the finite spectral resolution, constraining further sensitivity improvements. To address this, we consider a more general case where the signal field $\Omega_{\mathrm{s}}$ differs in frequency from $\Omega_{\mathrm{L}}$ by a detuning $\delta$. Under the rotating-wave approximation, the total MW driving field can be expressed as $\Omega_{\mathrm{L}} \rightarrow \Omega_{\mathrm{L}} + \Omega_{\mathrm{s}} e^{-i(\delta t + \phi)}$~\cite{Gordon2019,Jing2020a}, where $\phi$ denotes the relative phase between the signal and the local field.  Under this condition, the eigenvalues of the non-Hermitian Hamiltonian given in Eq.~(\ref{eq3}) become time-dependent due to the oscillating signal field,
\begin{equation}
	E_{\pm}(t) = \frac{-i \gamma_{\mathrm{c}} \pm \sqrt{4 (\Omega_{\mathrm{L}}^2+\Omega_{\mathrm{s}}^2)+8\Omega_{\mathrm{L}}\Omega_{\mathrm{s}}\cos{(\delta t+\phi)} - \gamma_{\mathrm{c}}^2}}{4}.
	\label{eq4}
\end{equation}
Therefore, by locking the laser frequency of the coupling field (set $\Delta_\mathrm{c}\simeq0$), the measurement of the signal MW field $\Omega_\mathrm{s}$ is converted into fluctuations in the EIT signal amplitude. This effectively transforms the frequency-based spectral measurement into an amplitude-based one, making the sensitivity independent of the spectral resolution. However, the amplitude variation in probe transmission induced by the same signal field $\Omega_\mathrm{s}$ strongly depends on the value of the local oscillator field $\Omega_\mathrm{L}$. 
\begin{figure}[tbp]
	\centering
	\includegraphics[width=\linewidth]{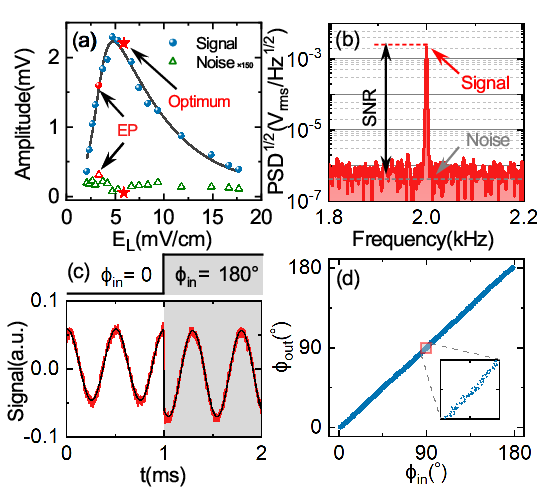}
	\caption{Performance of EP-enhanced electric field sensing. (a) Signal response amplitude {(blue dots) and noise (green triangles)} versus local MW field strength \(E_\mathrm{L}\) with fixed weak signal \(E_\mathrm{s}\). Maximum response (star) occurs slightly off the EP, optimizing signal-to-noise ratio. Blue spheres: experimental data; black line: fitting curve. (b) Power spectral density (PSD) of probe transmission at optimal point in (a), yielding electric field sensitivity of \(22.68(3)~\mathrm{nVcm^{-1}Hz^{-1/2}}\). (c) Time-domain probe transmission under phase-modulated \(\Omega_s\), showing optical signal phase inversion as MW input phase \(\phi_{\mathrm{in}}\) flips from 0 to \(180^\circ\) (shaded). (d) Measured output optical phase \(\phi_{\mathrm{out}}\) versus input MW phase \(\phi_{\mathrm{in}}\), demonstrating accurate, linear phase mapping. Inset: zoom-in highlighting phase resolution.
	}
	\label{fig4}
\end{figure}

{Figure}~\ref{fig3} { illustrates how the system responds to a weak signal field $\Omega_{\mathrm{s}}$ in the $\mathcal{PT}$-broken phase (Fig.}~\ref{fig3}(b1-d1)),{at the EP (Fig.}~\ref{fig3} {(c2-d2)) and $\mathcal{PT}$-symmetric phase (Fig.}~\ref{fig3}(b3-d3)). As shown in Fig.~\ref{fig3}(a), when the signal field $\Omega_\mathrm{s}$ is applied in the $\mathcal{PT}$-broken phase (labeled as (c1)), the real parts of the eigenvalues remain degenerate $\mathrm{Re}(E_\pm) = 0$ while the imaginary parts vary periodically in time $\mathrm{Im}(E_\pm)\propto\pm\Omega_{\mathrm{s}}\cos{(\delta t+\phi)}$, as illustrated in Fig.~\ref{fig3}(c1) (only one branch is shown for clarity). 
As shown in Fig.~\ref{fig3}(b1), this temporal modulation of the imaginary parts leads to linewidth oscillations of the eigenmodes (gray dashed lines), resulting in corresponding changes in the probe transmission amplitude (blue solid line). Consequently, in the spectral domain, a pronounced peak emerges at the signal frequency ($\delta/2\pi=2~\mathrm{kHz}$ in the experiment), as shown in Fig.~\ref{fig3}(d1), where both theoretical (top) and experimental (bottom) results are presented. In contrast, when the signal field $\Omega_\mathrm{s}$ is applied in the $\mathcal{PT}$-symmetry phase (labeled as (c3)  in Fig.~\ref{fig3}(a)), as shown in Fig.~\ref{fig3}(c3), the real parts of the eigenvalues exhibit time-dependent modulation $\mathrm{Re}(E_\pm)\propto\pm\Omega_{\mathrm{s}}\cos{(\delta t+\phi)}$ while the imaginary parts remain constant. As shown in Fig.~\ref{fig3}(b2), this causes the resonance positions of the eigenmodes (gray dashed lines) to shift, leading to corresponding changes in the probe transmission (orange solid line). In this regime, the signal manifests as a distinct spectral peak at its frequency $\delta/2\pi$, as demonstrated in Fig.~\ref{fig3}(d3) with both theoretical and experimental spectra.

Remarkably, when $\Omega_\mathrm{L} \simeq \Omega_\mathrm{EP}$, the signal field $\Omega_\mathrm{s}$ causes the system to continuously traverse the two phases (labeled as (c2) in Fig.~\ref{fig3}(a)), where both the real and imaginary parts of the eigenvalues oscillate nonlinearly, as shown in Fig.~\ref{fig3}(c2). Due to the nonlinear response near the EP, these oscillations deviate from simple sinusoidal behavior. Consequently, both the resonance shifts and linewidth modulations contribute to changes in the probe transmission spectrum, resulting in strong nonlinear features at multiple harmonics of the signal frequency, as illustrated in Fig.~\ref{fig3}(d2). Our theoretical predictions and experimental results are in good agreement, highlighting a new mechanism for optical field modulation based on EP-induced nonlinearity, and providing a versatile platform for exploring nonlinear dynamics.

We now examine the enhancement of electric field detection enabled by the nonlinear response near the EP. As shown in Fig.~\ref{fig4}(a), {the response amplitude and noise to a fixed signal field $E_\mathrm{s}$ vary with the local field $E_\mathrm{L}$. By optimizing the signal-to-noise ratio, the optimal point is reached not exactly at the EP but within its surrounding nonlinear regime.} This behavior is beneficial, as it avoids the eigenbasis collapse at the EP, thereby circumventing the excess fundamental noise associated with it \cite{Wiersig2020,Kononchuk2022}. These observations clearly demonstrate that the EP-induced nonlinearity can be harnessed to enhance the sensitivity of MW electric field detection. The optimal signal-to-noise ratio (SNR), marked by the red star, defines the operating point for highest sensitivity.To determine the system sensitivity, we measure the square root of the power spectral density (PSD$^{1/2}$) by feeding the time-domain output signals into a fast Fourier transform (FFT) spectrum analyzer. As shown in Fig.~\ref{fig4}(b), the optimal sensitivity reaches {$22.68(3)~\mathrm{nVcm^{-1}Hz^{-1/2}}$} under realistic experimental conditions ({see} \cite{supp1} {for details}). Moreover, the system remains capable of measuring the phase of the signal field. Figure~\ref{fig4}(c) shows the time-resolved probe transmission. When the signal field undergoes a $180^\circ$ phase flip via a phase shifter, the phase of the oscillatory probe signal correspondingly flips, as indicated by the shaded gray regions. By continuously scanning the input signal phase $\phi_\mathrm{in}$ using a phase shifter and reading out the corresponding output phase $\phi_\mathrm{out}$ with a lock-in amplifier, we realize continuous phase detection, as shown in Fig.~\ref{fig4}(d). This demonstrates the system’s ability to detect the phase of a MW field.

In summary, we demonstrate EP-enhanced Rydberg atomic sensing with both high sensitivity and phase resolution. By engineering dissipation to realize a second-order EP, we induce a nonlinear spectral response that amplifies weak MW signals. Near the EP, the AT splitting exhibits square-root scaling, yielding nearly 20-fold signal enhancement and achieve a sensitivity of {$22.68(3)~\mathrm{nVcm^{-1}Hz^{-1/2}}$} under realistic conditions. {The system enables the measurement of the amplitude, frequency, and phase of unknown microwave fields through the modulation of probe transmission near the EP. In combination with established polarization-retrieval methods}~\cite{Ang2023},{ the platform can in principle be extended to polarization-resolved microwave electrometry.} {In realistic environments, spatial inhomogeneity of microwave amplitude or polarization primarily introduces additional broadening, which may shift the EP position but does not alter the characteristic square-root response, ensuring robust enhancement. For multi-frequency fields, each frequency component generates a distinct peak in the probe-transmission spectrum, enabling straightforward frequency discrimination.} Our methodology opens up a versatile platform for exploring non-Hermitian physics in practical quantum metrology applications.
\begin{acknowledgments}
This work is supported by {National Key R\&D Program of China (Grant No. 2025YFA1411600)}, the National Natural Science Foundation of China (NSFC) ({Grants No.92476205}, No. 12361131576 and No. 92265205), and by Innovation
Program for Quantum Science and Technology (Grant No. 2021ZD0302104).
C. L. is supported by the China Postdoctoral Science Foundation (BX20230186) and
the Shuimu Tsinghua Scholar Program.
\end{acknowledgments}
%

\end{document}